# PARA UN ESTUDIO "COMPUTACIONAL" DE LOS INTELECTUALES *SATÉLITES*
# FOR A "COMPUTATIONAL" STUDY OF THE INTELLECTUALS SATELLITES

## Resumen

La presente propuesta extiende las reflexiones teóricas que llevamos sobre los "intelectuales satélites", un concepto que desea puntear un repertorio de intelectuales que la crítica no ha estudiado sobradamente y que emerge con la amplificación de los proyectos de digitalización de archivos y difusión en línea. Hoy, se necesita replantear y reevaluar los criterios con los que se solía definir, integrar o excluir obras y autores del/de un "canon" al mismo tiempo que creamos nuevos métodos para procesar estos datos.

## Abstract

The present proposal wishes to extend the theoretical reflections that we're currently working on: the "satellite intellectuals". This concept points a new "repertoire" of intellectuals that critics have not studied thoroughly and that is emerging with the amplification of the archives' digitalization of archives and their digital edition. Today, it is necessary to rethink and reevaluate the criteria we used to apply to define, integrate or exclude works and authors from, or in, a "canon". At the same time, it suggests new methods to process and deal with some concepts of computer tools and science.

## Palabras clave

Archivos, Digitalización, Literatura, Procesamiento de Datos.

## Fatiha Idmhand

Universidad de Poitiers
Institut des Textes et Manuscrits
Modernes UMR8132, Francia

Catedrática de Literatura y Humanidades Digitales en la Universidad de Poitiers e investigadora del equipo Archivos del Institut des textes et Manuscrits Modernes (UMR8132). Su investigación se centra en la literatura contemporánea, en sus procesos creativos y su circulación entre Europa (España, Francia) y el Río de la Plata (Uruguay) en contextos de conflictos y crisis y en Humanidades digitales.

FATIHA IDMHAND

# PARA UN ESTUDIO "COMPUTACIONAL" DE LOS INTELECTUALES *SATÉLITES*

*"El patrimonio escrito es la razón de ser de las bibliotecas y de los archivos, públicos y privados: en ellos se conservan las aventuras del pensamiento, las responsabilidades de la palabra dada, el tesoro del lenguaje y la memoria de las naciones."*
Pierre-Marc De Biasi, "El patrimonio escrito" [1].

*"Por lo tanto, el problema de las bibliotecas es doble: primero un problema de espacio y luego un problema de orden. (…): una cosa para cada lugar y cada lugar para su cosa y viceversa."*
Georges Pérec, Pensar, clasificar [2].

La presente propuesta desea extender las reflexiones teóricas que llevamos en la actualidad sobre los "intelectuales satélites", un concepto que desea puntear un repertorio de escritores e intelectuales que la crítica no ha estudiado suficientemente[3]. Se trata de autores/as que estuvieron relacionados/as con las figuras más ilustres del siglo xx y cuyo rol fue decisivo para la difusión de las obras e ideas de los más famosos. En nuestra opinión, su posicionamiento a la sombra de los más famosos, fue una vía para vehicular otro mensaje sobre sus compromisos, su grupo, su mundo y también sobre su época. Se trata, por ejemplo, de figuras como José Mora Guarnido o Jaime Sabartés, amigos íntimos de Federico García Lorca y Pablo Picasso[4] quienes obraron, desde la órbita, para construir las figuras públicas del poeta y del pintor y para difundir las ideas de sus grupos: *El Rinconcillo* y *Els Quatre Gats*. Sumamente activos en las redes de ambos artistas, a pesar de ser muy discretos, la crítica no los ha considerado adecuadamente hasta el día: o porque no encajaban con algunos criterios que conviene cuestionar hoy o, simplemente, porque los documentos, archivos y la información no se encontraba disponible hasta la actualidad[5]. Con la amplificación de los proyectos de digitalización de archivos y difusión en línea, vemos emerger cada vez más corpus en la web: requieren que replanteemos y reevaluemos los criterios con los que se solía definir, integrar o excluir obras y autores del/de un "canon". Podrá parecer contradictorio el hecho que propongamos una nueva categoría para situar a quienes fueron excluidos de otras categorías, no obstante, ésta parece oportuna de modo transitorio. Nos ayuda a identificar a autores que no figuran en el panorama habitual de la crítica y cuyo rol fue decisivo.

Nuestras primeras investigaciones sobre el tema están revelando que muchos de los "satélites" eligieron su posición de *testigos de primera fila* para difundir, desde otra vía, sus ideas utilizando, de algún modo, a la *estrella* del grupo. ¿Cuáles? Esta es una de las pistas que estamos ahondando en la actualidad reconstruyendo el itinerario biográfico de estas figuras. Para ello, indagamos también las posibilidades de trabajar, con la informática, y colectar datos e informaciones vinculados con la temática. Lo que se propone a continuación son primeras pistas prácticas para la construcción de los conjuntos de datos (*data sets*). Situamos esta propuesta desde el punto de vista de la elaboración de los datos porque es una etapa liminar fundamental para la difusión de los datos y también, para su comparación con otros datos que podrían ser colectados (*harvested*) en la web después. Estas propuestas resultan de experimentaciones realizadas entre el 2014 y 2017[6] que fueron probadas, como se explicará, sobre varios corpus de autores y que, según notamos, permiten mejorar sustancialmente la calidad de los datos brutos.

## 1. LOS INTELECTUALES *SATÉLITES*: ACTORES DE LAS "TRANSFERENCIAS CULTURALES"

El archivo del republicano español José Mora Guarnido (Alhama de Granada, España 1894-1968 Montevideo, Uruguay)[7], a partir del cual construimos gran parte de los experimentos aquí descritos, ilustra el mundo hispano del siglo xx (España y América Latina) marcado por conflictos mortales, dictaduras y grandes desplazamientos de población. Desde su exilio en Montevideo, Mora ha producido todo tipo de textos para denunciar el autoritarismo, las guerras y el exilio político; es cierto que conoció y vivió sucesivamente: la dictadura de Primo de Rivera (1923-1931), la primera crisis financiera (1929), la dictadura de Gabriel Terra en el Uruguay donde se había exiliado (1931-1938), la Segunda (y breve) República Española (1931-1936), la Guerra Civil (1936-1939) y la dictadura de Franco cuyo final no conoció (1939-1975), la Segunda Guerra Mundial (1939-1945), la Revolución Cubana (1953-1959) y la repercusión de la extensión de las ideas castristas por América latina desde el año 1956.

La lectura cruzada de sus textos de ficción, artículos de prensa y correspondencia permite medir la importancia de su compromiso político y cultural, y la de gran parte de los intelectuales de su generación. Numerosos son los elementos, presentes en el archivo, que revelan las conexiones de José Mora Guarnido con las redes intelectuales más prestigiosas de la época y que permiten entender la manera como éstas se conectaban a través del mundo durante conflictos que generaron incesantes movimientos de personas[8]. Entre ellos, las 432 cartas que componen su epistolario[9] son las que mejor hacen revivir los diferentes estados del mundo entre 1923 y 1960: se trata de conversaciones epistolares mantenidas con la familia que siguió viviendo en Granada y con los amigos conocidos en distintos "mundos": aquellos con quienes fundó el *Rinconcillo del café Alameda* (como Manuel de Falla y Melchor Fernández Almagro); los que conoció en Madrid (en el Ateneo o en la Residencia de estudiantes o en el Café Pombo como Ramón Gómez de la Serna, Guillermo de Torre o Gonzalo Losada); o aquellos con los que estuvo relacionado mientras fue Cónsul de la República Española en el Uruguay (como los políticos Enrique Diez Canedo o Rodrigo Soriano[10]). Los intercambios con Jaime Sabartés son ejemplares en este sentido.

José Mora Guarnido le conoció durante su estadía prolongada en el Uruguay a principios de los treinta. Antes, Jaime Sabartés había pasado unos años en Guatemala. Sus trayectorias dejan entrever algunas analogías[11], en particular, sus vidas cerca de dos amigos que se convirtieron en "monstruos" de la cultura del siglo xx y la concordancia entre sus dos "galaxias" artísticas y culturales. Por eso, pensamos que indagar ambas redes permite entender dos de las más importantes del mundo hispánico de la primera mitad del siglo xx. Si se extiende un poco más la reflexión a otros corresponsales de José Mora Guarnido, también encontramos vinculados con él los nombres de Gonzalo Losada, Guillermo De Torre, Enrique Amorím o Enrique Díez Canedo. Seguir estas pistas nos abre la puerta a conexiones más importantes con otras redes del mundo latinoamericano de los años 1950, en el Río de la Plata en particular. Esta pista de trabajo nos interesa desde la perspectiva de del estudio de las *transferencias culturales*. El concepto, forjado por Michel Espagne en los años 1980[12], propone estudiar los procesos de "re-semantización" de los objetos culturales en las transferencias culturales en una perspectiva dinámica que analiza la reinterpretación semántica de los objetos culturales después de cada traspaso, tomando en cuenta el rol de "actores" de las transferencias, los que Michel Espagne llama *mediadores.* Son, según él, los exiliados, traductores, editores, periodistas o académicos que facilitan la circulación y transmisión de los objetos culturales. Desde esta noción, el intelectual *satélites* se revela un caso particular de *mediador* al participar, desde *la primera fila*, en la difusión y reconstrucción de aquellos objetos culturales. Creemos que sus manuscritos y archivos no sólo revelan el papel decisivo que tuvieron como operadores claves de las transferencias culturales, sino que también cuentan desde otra perspectiva, la historia de la circulación de ideas.

**2. PENSAR/CLASIFICAR**

Planteado el contexto del estudio, surge la pregunta de cómo cartografiar e ilustrar el entramado de estas redes. Preguntándonos sobre qué tipos de grafos convendría usar para representar esta red intelectual y social culturalmente muy activa y cuya importancia radica, no sólo en los distintos nodos sino también en los enlaces y flujos entre los nodos, nos dimos cuenta, rápido, de la importancia de tener datos "procesables". Las herramientas (*software*) que existen en la actualidad y que permiten construir visualizaciones requieren, antes de todo, datos pulidos. Los proyectos en Humanidades Digitales, suelen descuidarlos lo cual engendra varios procesos de "curación" cada vez que se desea usar una herra-mienta distinta y confrontar hipótesis. De modo que, y en relación con el proyecto científico, llegamos a preguntarnos sobre la manera de construir nuestros datos brutos, con calidad, y con el fin de procesarlos fácilmente simplificando su gestión. Dicho de otro modo, si muchas plataformas ofrecen, en la actualidad, interfaces que facilitan el trabajo de los *humanistas digitales* al ayudarles facilitar la introducción de datos, es de notar que a la hora de extraerlos, para analizarlos o procesarlos en otra herramienta, entonces es cuando aparecen las dificultades. Pensamos que parte de las carencias en la explotación de los datos de las humanidades se deben a esta falta de flexibilidad. Nuestras distintas experiencias con algunas herramientas "WYSIWYG" o con algunos CMS como Omeka[13] nos mostraron que usar un programa informático disponible en el mercado para adaptarlo a nuevas necesidades y facilitar la integración de los datos, sí simplifica la vida al principio, pero puede complicarla en el momento en el que la extracción es necesaria para realizar análisis. Simplemente porque a veces, el formato de salida (*output*) requiere la intervención de alguien para procesarlo. O sea que lo que parece sencillo al inicio (la integración de datos) se vuelve complejo cuando se trata de realizar los análisis, de cruzar y comparar los datos. ¿Qué pasa cuando no hay ingeniero o colega siempre disponible? Nuestra propuesta sugiere salir de la dependencia a la herramienta en el momento de la construcción de los datos y reconsiderar, antes que nada, la forma y calidad de los datos brutos. Al haber tenido la oportunidad de organizar un archivo, el de José Mora Guarnido (se encontraba custodiado por la familia a quien Mora lo había legado y no en instituciones públicas o privadas como bibliotecas, archivos o fundaciones), pudimos construir un modelo para la integración de datos. Se trata de un esquema de metadatos que toma en cuenta:

- la gran variedad del archivo físico (manuscritos, tapuscritos, correspondencias, fotografías, grabaciones audio, fotografías, dibujos, etc.)

- la necesaria calidad de los datos brutos
- los desafíos de las investigaciones científicas antes planteadas y que conciernen la reevaluación del rol de los *satélites* en las redes intelectuales del exilio del siglo XX y su contribución, desde la primera fila, a la difusión y a la construcción de nuevas corrientes culturales y artísticas.

Para ello, se consideró que los datos elaborados, en relación con el proyecto científico, también debían facilitar la "minería de datos" (data mining). Los datos describen "una cosa para cada lugar y cada lugar para su cosa y viceversa"[14]: así, los más de 40 campos de metadatos del esquema informan sobre la forma del manuscrito (cuaderno, folio, etc.), su tipo, su naturaleza (grabación, iconografía, prensa, correspondencia, etc.), su fecha, etc. y sobre el estado "genético"[15] del documento, o sea, la etapa en la cual se sitúa en el proceso de creación de la obra. Esta información abre la posibilidad de una búsqueda por proyecto, y no solo por documento: se indica por lo tanto si el documento es, o no, un borrador de otro manuscrito, si forma parte o no, de un proyecto y a qué fase del proyecto corresponde. Esta información se ha traducido con la creación de un nuevo descriptor el "Dosier genético", y traducido en un campo específico llamado "Relaciones genéticas". Allí se identifica, a partir del código del documento, su estado genético: C1 para Copia 1, C2 para Copia 2, etc. La etiqueta "Relaciones genéticas" se inspira de lo que ya existe en algunas normas internacionales como el Dublin Core, que tiene una etiqueta "Relation". Para vincular los documentos, la signatura es importante relacionar la fuente material y su artefacto digital. Para ello, dimos la misma signatura a los dos y para crearla, organizamos un Plan de clasificación flexible.

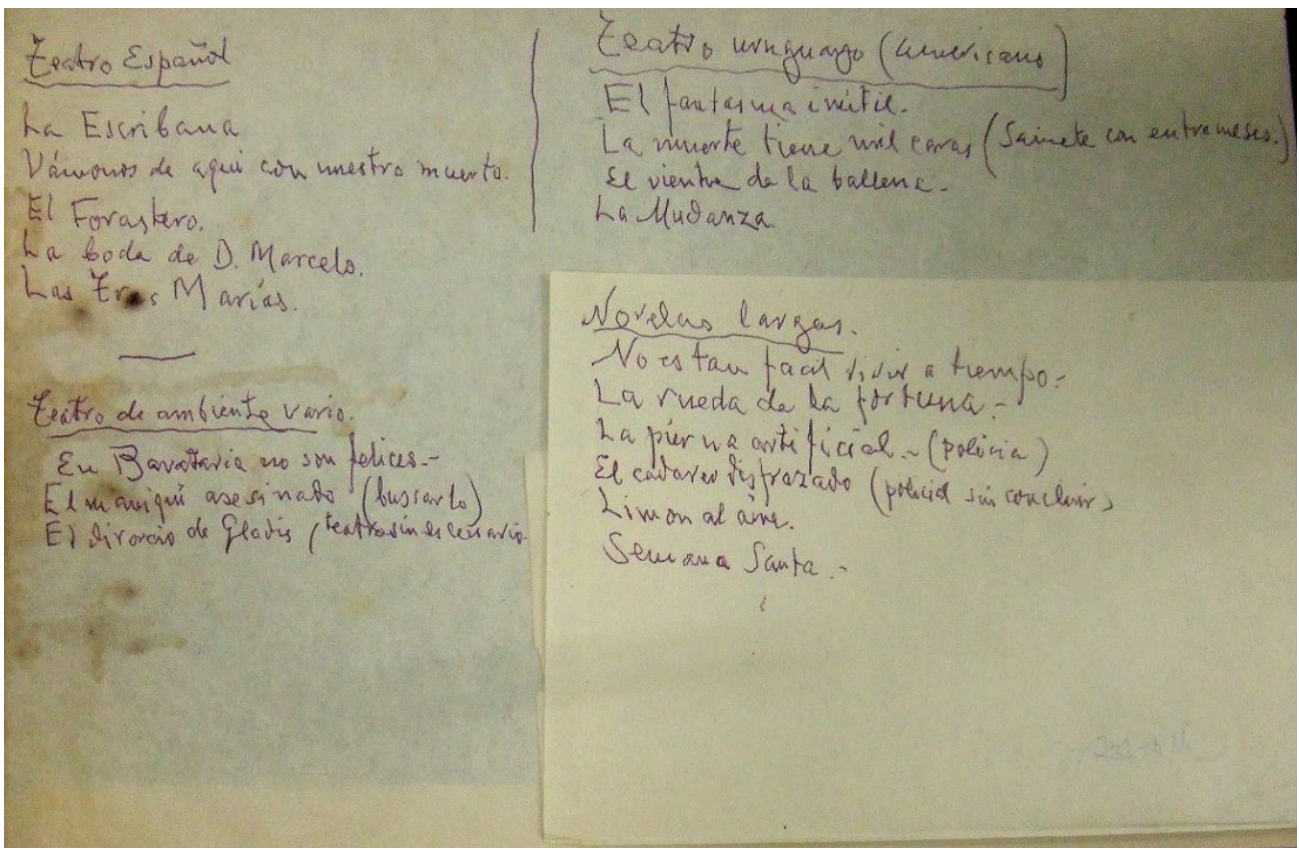

*Fig. 1. Inventario de José Mora Guarnido | JMG-EA-01 @Archivo José Mora Guarnido.*

El Plan de clasificación organiza y ubica, física e intelectualmente, los documentos del archivo[16]: es un ordenamiento que da sentido a cada objeto del archivo. En principio, se realiza en función de una serie de criterios como la cronología, la geografía, el tema o las fechas, y en un servicio de archivo o en una biblioteca. Cuando es así, respeta el "patrón" predeterminado por la institución. Éste relaciona el organismo, el archivo, su organización intelectual y física, con el público que lo va a consultar. El plan de clasificación es, por lo tanto, es el objeto por el cual se construye la mediación entre el archivo y el público[17] y es único porque cada archivo y cada autor es único: se elabora de acuerdo con las especificidades de cada artista.

¿Cómo conciliar la especificidad del archivo con un esquema de metadatos generalista, de cara a un procesamiento automático y a una mayor interoperabilidad de los datos? Normalmente, orden intelectual y material coinciden en una signatura única: la designación es por lo tanto un momento crucial. A la luz de todas estas reglas, construimos un modelo de plan práctico y "extensible" al ser aplicable a cualquier tipo de proyecto. Este plan de clasificación tiene cinco categorías principales ordenadas de A a E; a partir de estas grandes categorías, se crean subcategorías que reflejan las particularidades temáticas y contextuales de cada autor:

   A. Obras, creaciones, producciones
   B. Correspondencia
   C. Documentación
   D. Archivo biográfico
   E. Recepción

La signatura consta entonces de una numeración simple que mezcla letras y números y que asocia documento físico y digital. La signatura es alfanumérica, secuencial y/o cronológica, en función de la naturaleza del documento al que se refiere y su categoría. En comparación con lo que se practica en las instituciones (bibliotecas, archivos, etc.), es un sistema que no brinda ninguna información sobre el establecimiento que "hospeda" el archivo: la información aparece entre los metadatos, en una categoría llamada "Localización" (*source*). Así, por ejemplo, se describe la correspondencia de José Mora Guarnido (JMG) en la categoría B con una signatura que integra la fecha de las cartas con una letra a, b, c, etc. para distinguir dos cartas que tendrían las mismas fechas:

   A. Obras, creaciones, producciones
   B. Correspondencia
      JMG-B-1923-00-00
      JMG-B-17-08-1923
      JMG-B-09-06-1923
      JMG-B-1924-09-17a
      JMG-B-1924-09-17b

La originalidad de la propuesta radica en la creación de un sistema de clasificación independiente del establecimiento: permite que se conserve el mismo Plan si el archivo es trasladado de una estructura a otra. Probando el sistema con otro archivo, el del dramaturgo uruguayo Carlos Denis Molina (Uruguay, 1916-1983)[19], usamos la misma metodología para categorizar sus obras de teatro, relatos, artículos de prensa, etc.:

   A. Obras, creaciones, producciones
      AA. Teatro
         AA1. Interludios
            01. Celebración en la plaza [C1]
   B. Correspondencia
   C. Documentación
   D. Archivo biográfico
   E. Recepción

CDM indica a qué fondo pertenece el documento (Carlos Denis Molina), AA1 indica que el documento está en la categoría *A. Obras, creación, producciones*, y en la serie *AA. Teatro*, en la subcategoría *AA1. Interludios*. Las subcate-gorías del plan de clasificación ilustran la especificidad de este autor y hacen hincapié en la vertiente teatral de su producción artística. *01* indica que es el primer documento de la serie y *C1* que este es el primer estado genético del proyecto de creación *Celebración en la plaza [Copia 1]*. Las "Relaciones genéticas" del documento CDM-AA1-01-C1 son CDM-AA1-01-C2 y CDM-AA1-01-C3.

**3. NORMALIZAR PARA PROCESAR Y EXPLOTAR**

La organización del archivo, la creación de las signaturas y la elaboración simultánea de su descripción informática supuso pensar, también, en la forma de construir, estructurar y categorizar la información en bruto. Se optó por almacenarla en un archivo de texto en formato CSV[20] para una mayor flexibilidad[21]. Al usar un CSV, el objetivo era facilitar el uso de todo tipo de herramientas de "minería de datos": en lugar de usar un único *software* y tener que adaptarlo en función de distintas necesidades, se privilegió el formato bruto CSV al inicio de la preparación de los datos y la normalización de éstos en función de las recomendaciones habituales: presentación de los nombres, apellidos, nombres de lugares, países, fechas, respecto de los estándares (ISO), etc.

Luego, a partir del CSV, y en función del objetivo, ya sea el uso de un CMS para presentar los datos o para publicarlos en un sitio web, la creación de un fichero Dublin Core, XML-EAD o incluso de *headers* para ficheros en XML-TEI[22], sólo falta transformar el fichero bruto y sus etiquetas en función de lo que se precise. La metodología descrita se inspira en el *single-source publishing*[23], es un procedimiento que permite administrar los contenidos a partir de una única fuente. A partir del CSV, se puede presentar los datos de diferentes formas, en distintos medios, y esto más de una vez. Es una técnica que se revela poco costosa: se desarrollan, cuando falta hace, pequeños programas informáticos (*software*) para transformar los datos brutos. De modo que el trabajo más laborioso y costoso se lleva a cabo al inicio del proyecto, o sea, cuando el científico prepara sus datos: se supone que un trabajo conceptual que anticipe, antes de que se construyan las categorías y se formalicen los contenidos, la forma de la información y la manera como se explotará[24].

Esta misma metodología resulta aplicable en el caso de la recolección de datos de otros proyectos[25]. En esta caso, el *harvesting* es realizado por un programa (algoritmo) que se realiza específicamente para cosechar datos, éstos se recogen en el csv siendo adaptados al esquema relacionado con el proyecto científico a través de un *mapping.* Después, se depuran los datos y se normaliza su presentación para poder, luego, compararlos y analizarlos. La ventaja del CSV es que numerosas herramientas lo usan, de entrada; así, la refinación y normalización puede ser realizada con OpenRefine que devuelve un CSV al final del proceso, también se puede experimentar cálculos estadísticos con "R" o construir visualizaciones con herramientas como Gephi o Cytoscape. Todas estas herramientas (actualmente de moda) son útiles para apoyar las problemáticas científicas pero para ir de una a otra, siempre hace falta volver a un formato flexible.

En el caso de nuestro estudio sobre intelectuales *satélites*, pudimos notar que algunas informaciones se revelan particularmente operantes a la hora de construir una perspectiva "distante" tal como fue teorizada por Franco Moretti en su libro *Distant Reading*[26] sobre estos nuevos corpus. Estos datos son los que integramos en el CSV y entre ellos, consideramos importantes las informaciones sobre los lugares, fechas de nacimiento y muerte de los intelectuales para cartografiar el itinerario de su exilio; su profesión, para circundar mejor la figura del *mediador de las transferencias culturales* y sus acciones (publicaciones o creaciones: fecha, lugar y naturaleza o tipo) y para relacionar el objeto cultural

10

creado con los "objetos transferidos". También integramos el nombre de las figuras que acompañaron, las "estrellas". Para la descripción de la correspondencia, se integran informaciones sobre el remitente como el lugar de envío de la carta. En la actualidad, y a partir de los datos que llegamos a almacenar, vemos que los intelectuales *satélites* podrían haber actuado como "correa de transmisión" entre las diferentes redes. Su inserción en ellas se debe al hecho que también eran artistas, creadores de contenidos. A pesar de ser *testigos de primera fila*, no pueden ser considerados solamente como enlaces: conviene encontrar recursos o medidas para hacer hincapié en el hecho que también eran artistas; una vía posible podría ser comparando la naturaleza de las informaciones o solicitudes que reciben, con la que transmiten y con las realizaciones artísticas y/o culturales de ellos, y de las *estrellas*. Es un experimento que estamos llevando a cabo en la actualidad inspirándonos en algunas tecnologías de *text mining*.

En todo caso, terminaremos recalcando el proceso intelectual que acompaña nuestra indagación y cuya meta es tratar un corpus de autores emergente en la web, poco considerado por la crítica hasta el día, integrando las posibilidades de procesar, automáticamente, esta nueva información. El objetivo nuestro es sacar nuevas conclusiones y conocimientos a partir de los archivos que están saliendo hoy. Al mismo tiempo, vemos que se requiere una nueva epistemología: la posibilidad de recolectar datos parece ser infinita ¿cómo se van a investigar?, ¿a comparar? y ¿a partir de qué criterios? En este sentido, como se ha dicho en la introducción, conviene repensar muchos de los criterios habituales de las ciencias de los textos porque algunos resultan "perecidos" en este contexto.

## NOTAS

[1] DE BIASI, Pierre-Marc. "Le patrimoine écrit". En línea: http://www.item.ens.fr . [Fecha de acceso: 25/05/2018].

[2] PEREC, Georges, *Penser/Classer*. París: Éditions du Seuil, 2003 (primera ed. 1985), pág. 38.

[3] IDMHAND, Fatiha, CASACUBERTA ROCAROLS, Margarita. "Intelectuales "satélites". Hacia un nuevo enfoque sobre la circulación de la literatura y de la cultura", 19 (2017), http://revistaseug.ugr.es/index.php/letral. [Fecha de acceso: 15/05/2018].

[4] Ver IDMHAND, Fatiha, CASACUBERTA ROCAROLS, Margarita. "Jaume / Jaime / Jacobus Sabartés: la biografía como autorretrato". En: Congreso Picasso i identitat, 2017, http://museupicassobcn.org/congres-internacional/casacuberta-idmhand/. Fecha de acceso: 23/04/2018].

[5] El libro *Figuras del 36*, ofrece una muestra inédita de investigaciones sobre figuras poco o mal conocidas del exilio español del 36. Ver IDMHAND, Fatiha, CASACUBERTA ROCAROLS, Margarita, AZNAR SOLER, Manuel, DEMASI, Carlos. *Figuras del 36*. Bruselas: Editorial PIE Peter Lang Coll, prevista para el 2018.

[6] Parte de estas propuestas son el resultado del proyecto CHISPA que fue financiado por la Agence Nationale de la Recherche de 2014 a 2017 : http://www.agence-nationale-recherche.fr/Project-ANR-13-JSH3-0006. Ver http://chispa.hypotheses.org/. [Fecha de acceso: 15/04/2018].

[7] http://guarnido.nakalona.fr/. [Fecha de acceso: 15/04/2018].

[8] Algunos objetos y regalos conservados en el archivo también son indicios de otras relaciones de amistad que no dejaron huella epistolar, pero sí existieron. Es el caso de Margarita Xirgú, también radicada en el Uruguay desde la Guerra civil española, que regaló a la pareja José Mora Guarnido y Esther Morales una foto con una afectuosa dedicatoria que revela el contacto entre las parejas. Pasa lo mismo con esa "servilleta" de bar que le fue mandada como postal desde Roma, firmada por los "amigos" Rafael Alberti (España, 1902-1999) y Toño Salazar (El Salvador, 1897-1986). Ver: https://guarnido.nakalona.fr/items/show/1528. [Fecha de acceso: 25/03/2018].

[9]BASSO, Eleonora, DEMASI, Carlos, DEI CAS, Norah Giraldi, IDMHAND, Fatiha. "Trayectoria de José Mora Guarnido. Espejo de un intelectual entre España y América (1923-1939)". En: DE MORA, Carmen, GARCÍA MORALES, Alfonso (Eds.). *Viajeros, diplomáticos y exiliados. Escritores hispanoamericanos en España (1914-1939)*. Bern, Berlin, Bruxelles, Frankfurt am Main, New York, Oxford, Wien: P.I.E. PETER LANG, 2012, Tomo II, págs. 517-539. [en línea en HAL]. [Fecha de acceso: 13/05/2018].

[10]https://guarnido.nakalona.fr/items/browse?search=&advanced=correspondances. [Fecha de acceso : 26/04/2018].

[11]Citamos las "estrellas" Lorca/Picasso, los grupos a los que pertenecían El Rinconcillo/Els Qautre Gats, el rol de las ciudades de Granada/Barcelona en la emergencia de sus ideas vanguardistas, el compromiso político y el Exilios, la escritura de Biografías como género privilegiado para inscribir la figura propia en la sombra de la estrella, etc. Ver IDMHAND, Fatiha, CASACUBERTA ROCAROLS, Margarita. "Jaume / Jaime / Jacobus Sabartés: la biografía como autorretrato". En: Congreso Picasso i identitat, 2017, http://museupicassobcn.org/congres-internacional/casacuberta-idmhand/. Fecha de consulta : 18/05/2018].

[12]ESPAGNE, Michel. *Les transferts culturels franco-allemands*. París: PUF, 1999.

[13]Ver: http://chispa.hypotheses.org/category/on-a-teste-pour-vous . [Fecha de acceso: 15/03/2018].

[14]PEREC, Georges, *Penser/Classer…* Op. cit., pág. 38.

[15]La crítica genética, desarrollada desde finales de los 1960, analiza los procesos de creación de las obras desde sus apuntes liminares, borradores y materiales textuales. http://www.item.ens.fr/thematique/ . [Fecha de acceso : 15/03/2018].

[16]Jérôme Pouchol Mettre en œuvre un plan de classement », Villeurbanne, Presses de l'Enssib, coll. « La Boîte à outils », vol. 18, 2009. En línea: http://bbf.enssib.fr/, in Bulletin des bibliothèques de France, nº 2, 2010.

[17]Hasta el 2010, no existía, al menos en Francia, un modelo de "plan de clasificación" para los archivos de escritores contemporáneos, sobre cuando eran custodiados por privados. Fue en 2010 cuando la Biblioteca nacional de Francia publicó una propuesta, el plan para la "Descripción de manuscritos y registros de bibliotecas modernas y contemporáneas-DeMArch" Ver: http://www.bnf.fr/fr/professionnels/normes_catalogage_francaises/a.ead_demarch.html . [Fecha de acceso: 25/03/2018].

[18]Ver: https://guarnido.nakalona.fr/items/show/2106. [Fecha de acceso: 25/03/2018]. José Mora Guarnido intentó organizar su archivo al final de su vida. Nos dejó indicios sobre la particularidad de su creación ficcional, rigurosamente separada entre dos mundos: España/América, como su vida, partida por el exilio.

[19]Ver: https://molina.nakalona.fr . También se probó el mismo sistema con el archivo de Felisberto Hernández https://hernandez.nakalona.fr, el de Carlos Liscano https://liscano.nakalona.fr y el de Fernando Aínsa https://ainsa.nakalona.fr. [Fecha de acceso: 25/03/2018].

[20]El CSV (comma-separated values) es un formato abierto, sencillo y ampliamente compatible.

[21]Se puede descargar el esquema así como todas las herramientas descritas en este trabajo en https://github.com/ANR-CHispa . [Fecha de acceso: 15/03/2018].

[22]De momento, cuando las escasas transcripciones de textos que realizamos fueron guardadas en ficheros .txt. No alcanzamos indagar, por ahora, los XML TEI. Sin embargo, nuestra metodología facilita la creación de *headers TEI* y se basa en la versión P5 de la TEI.

[23]Ver: https://en.wikipedia.org/wiki/Single-source_publishing . [Fecha de acceso: 25/03/2018].

[24]Ver por ejemplo la manera cómo, a partir del esquema de metadatos en CSV, se pudo usar Omeka en: http://guarnido.nakalona.fr/ por ejemplo. Este sitio web sólo "presenta los datos", los *exhibe* pero Omeka no fue usado para inetgrarlos. Ver también los ejemplos: https://molina.nakalona.fr; https://hernandez.nakalona.fr; https://liscano.nakalona.fr o https://ainsa.nakalona.fr. [Fecha de acceso: 25/03/2018]. Nakalona es servicio de la infraestructura HumaNum (https://www.huma-num.fr/services-et-outils) que ofrece a los proyectos científicos un sitio web (nakalona.fr), y sobre todo un servidor donde se puede almacenar los datos brutos. [Fecha de acceso: 25/03/2018].

[25]Usando herramientas como Omescrap por ejemplo, para extraer, en un CSV, datos de otros Omekas https://github.com/ANR-CHispa. [Fecha de acceso: 15/05/2018].

[26]MORETTI, Franco. *Distant reading*. London/New-York: Verso, 2013.